\documentclass[journal,comsoc]{IEEEtran}
\usepackage[T1]{fontenc}

\ifCLASSINFOpdf
\else
\fi
\usepackage[cmintegrals]{newtxmath}

\usepackage{cite}
\usepackage{graphicx}
\usepackage{amsmath}
\usepackage{caption}
\usepackage{subcaption}

\begin{document}
%

\title{Non-Saturated Throughput Analysis of\\ Coexistence of Wi-Fi and Cellular With Listen-Before-Talk in Unlicensed Spectrum}

%
%
%

\author{Yongjae~Kim,~Yujae~Song,~Yonghoon~Choi,\IEEEmembership{~Senior~Member,~IEEE},
        and~Youngnam~Han,\IEEEmembership{~Senior~Member,~IEEE}
\thanks{This work was supported by 'The Cross-Ministry Giga KOREA Project' grant from the Ministry of Science, ICT and Future Planning, Korea.}
\thanks{Copyright (c) 2015 IEEE. Personal use of this material is permitted. However, permission to use this material for any other purposes must be obtained from the IEEE by sending a request to pubs-permissions@ieee.org.}
\thanks{Y. Kim, and Y. Han are with the Department
of Electrical Engineering, Korea Advanced Institute of Science and Technology, Daejeon, Korea (e-mail: \{yongjaekim, ynhan\}@kaist.ac.kr).}
\thanks{Y. Song is with ICT R\&D Unit, Korea Institute of Ocean Science and Technology, Gyeonggi-do, Korea (e-mail: yjsong@kiost.ac.kr).}
\thanks{Y. Choi is with the Department of Electrical Engineering, Chonnam National University, Gwangju, Korea (e-mail: yh.choi@jnu.ac.kr).}
}

%
%

\markboth{IEEE TRANSACTIONS ON VEHICULAR TECHNOLOGY}%
{Shell \MakeLowercase{\textit{et al.}}: Bare Demo of IEEEtran.cls for IEEE Communications Society Journals}
%



\maketitle

\begin{abstract}
This paper analyzes the coexistence performance of Wi-Fi and cellular networks conditioned on non-saturated traffic in the unlicensed spectrum.
Under the condition, the time-domain behavior of a cellular small-cell base station (SCBS) with a listen-before-talk (LBT) procedure is modeled as a Markov chain, and it is combined with a Markov chain which describes the time-domain behavior of a Wi-Fi access point.
Using the proposed model, this study finds the optimal contention window size of cellular SCBSs in which total throughput of both networks is maximized while satisfying the required throughput of each network, under the given traffic densities of both networks.
This will serve as a guideline for cellular operators with respect to performing LBT at cellular SCBSs according to the changes of traffic volumes of both networks over time.

\end{abstract}

\begin{IEEEkeywords}
Coexistence, licensed-assisted access (LAA), non-saturated traffic, unlicensed spectrum, Wi-Fi.
\end{IEEEkeywords}

%
\IEEEpeerreviewmaketitle

\section{Introduction}
%
%
%
%
In recent years, enabling cellular small-cells to operate in the unlicensed spectrum at typically 5GHz has received considerable attention as one of the solutions to cope with a spectrum scarcity problem. 
Since the unlicensed spectrum is available to anyone, it can be used fronthaul links as well as access links in 5G ultra dense networks \cite{Haijun2016}.

There are several types of LTE and Wi-Fi coexistence models in sharing the unlicensed spectrum: LTE-unlicensed (LTE-U), licensed-assisted access (LAA), and MulteFire \cite{Zhang2015, Abinader2014, Almeida2013, Song2016, multefire, Song2015, Chen2015, 3gpp, ericsson}. 
LTE-U with no regulatory requirement for listen-before-talk (LBT) is based on an adaptive on/off switching of cellular small-cells \cite{Zhang2015, Abinader2014, Almeida2013, Song2016}. On the contrary, both LAA and MulteFire abide by a listen-before-talk (LBT) procedure introduced in 3GPP Release 13 to access the unlicensed spectrum \cite{multefire, Song2015, Chen2015, 3gpp, ericsson}. The difference between LAA and MulteFire is whether an anchor in the licensed spectrum is required \cite{ multefire}. Hereafter, we focus on LAA-based medium access mechanisms. 

In \cite{Song2015}, the authors propose a Markov chain that describes the behavior of a cellular small-cell base station (SCBS) with a LBT procedure in the unlicensed spectrum, and then present the coexistence performance of Wi-Fi and cellular networks with different LBT procedures. The analysis in \cite{Song2015} is performed under the condition that Wi-Fi access points (APs) and cellular SCBSs always have packets to transmit, i.e., a saturated traffic condition. In \cite{Chen2015}, an analytic model which evaluates the coexistence performance between cellular and Wi-Fi networks under the non-saturated traffic condition is presented for the first time. The authors show the validity of adopting a LBT procedure in a Wi-Fi and cellular coexistence scenario by comparing with the case that the LBT procedure is not adopted. However, to the best of our knowledge, an analysis with respect to the effect of adjustment in the LBT parameter of cellular SCBSs, i.e., a contention window (CW) size, under the non-saturated traffic condition on the coexistence performance has not been identified in the literature before.

The main contributions of this paper are listed as follows:
\begin{itemize}
\item We propose an analytical model, i.e., a Markov chain, that describes the time-domain behavior of a cellular SCBS under the non-saturated traffic condition with the LBT procedure described in 3GPP TR 36.889 \cite{3gpp}, and it is combined with a Markov chain describing the time-domain behavior of a Wi-Fi AP introduced in \cite{Malone2007}.
\item Based on the analytical model, we investigate the throughputs of Wi-Fi and cellular nodes according to the change in traffic densities of both networks. For evaluating the coexistence performance, we adopt the concept of graceful coexistence, which is defined as the condition that the throughput of each node under a scenario with ${n_W}$ Wi-Fi and ${n_C}$ cellular nodes is better than that of each node under a scenario with homogeneously deployed ${n_W+n_C}$ Wi-Fi APs \cite{Song2015}. With the definition of graceful coexistence, we identify whether the graceful coexistence is satisfied by adjusting the CW size of the cellular SCBSs in all traffic densities of both networks. Then, we find the optimal CW size, by which total throughputs of both networks is maximized while satisfying the graceful coexistence.
\item Our results will serve as a guideline for cellular operators in performing LBT at cellular SCBSs according to the changes in traffic volumes of both networks over time, to coexist well with Wi-Fi APs in the unlicensed spectrum.
\end{itemize}


\section{Coexistence Performance Analysis}
We consider a scenario in which ${n_W}$ Wi-Fi APs and ${n_C}$ cellular SCBSs coexist on the same channel in the 5GHz unlicensed spectrum and operate under the non-saturated traffic condition. In \cite{Bianchi2000}, a Markov chain model is proposed to investigate the saturation throughput performance of the 802.11 distributed coordination function. As the extension of \cite{Bianchi2000}, \cite{Malone2007} provides a Markov chain for describing the time-domain behavior of a Wi-Fi AP under the non-saturated traffic condition. In this work, we propose a Markov chain to describe the time-domain behavior of a cellular SCBS under the non-saturated traffic condition in the unlicensed spectrum, and it is combined with the Markov chain of \cite{Malone2007} for evaluating the coexistence performance between cellular and Wi-Fi networks. Same as \cite{Song2015, Malone2007, Bianchi2000}, a fundamental assumption is that Wi-Fi and cellular nodes have a fixed collision probability regardless of their previous transmission history.

\subsection{Wi-Fi AP Model}
In \cite{Malone2007}, the Markov chain consists of post-backoff and backoff stages. The post-backoff stage stands for a set of states presenting the time-domain behavior of a Wi-Fi AP in case of not having an additional packet to transmit after successful packet transmission. Whereas, the backoff stage stands for a set of states presenting that in case of having a packet to transmit. In the post-backoff stage, there are states ${(0,k)_e}$ for $k \in [0,{W_0} - 1]$, where $k$ denotes the backoff counter and ${W_0}$ denotes the minimum CW size of the Wi-Fi AP. In the ${i}$-th backoff stage, there are states ${(i,k)}$ for $i \in \left[ {0,m} \right]$ and $k \in [0, {W_i} - 1]$, where ${m}$ denotes the maximum backoff stage and ${W_i}$ denotes the CW size in the ${i}$-th backoff stage, ${W_i} = {2^i}{W_0}$. The backoff counter ${k}$ is uniformly chosen in the range of $\left[ {0,{W_i} - 1} \right]$. It is decreased by 1 when the channel is sensed to be idle. In state ${(0,k)_e}$, the state transition occurs to state ${(0,k-1)}$ when a packet arrives at the Wi-Fi AP and the channel is sensed to be idle. The transmission of a packet is attempted at ${k=0}$. The backoff stage ${i}$ increases by 1 if a transmission attempt results in a collision, otherwise it is reset to 0. 

Let $b_{{{\left( {0,0} \right)}_e}}^W$ denote the stationary probability of state ${{{\left( {0,0} \right)}_e}}$. It can be obtained by using the normalization condition as follows \cite{Malone2007}:
\begin{equation}
\begin{array}{l}
b_{{{\left( {0,0} \right)}_e}}^W = \left[ {\left( {1 - {q_W}} \right) + \frac{{q_W^2{W_0}\left( {{W_0} + 1} \right)}}{{2\left( {1 - {{\left( {1 - {q_W}} \right)}^{{W_0}}}} \right)}} + \frac{{{q_W}\left( {{W_0} + 1} \right)}}{{2\left( {1 - {q_W}} \right)}}} \right.\\
 \cdot \left( {\frac{{q_W^2{W_0}}}{{1 - {{\left( {1 - {q_W}} \right)}^{{W_0}}}}} + \left( {1 - P_{idle}^W} \right)\left( {1 - {q_W}} \right) - {q_W}P_{idle}^W\left( {1 - {p_W}} \right)} \right)\\
 + \frac{{{p_W}q_W^2}}{{2\left( {1 - {q_W}} \right)\left( {1 - {p_W}} \right)}} \cdot \left( {\frac{{{W_0}}}{{1 - {{\left( {1 - {q_W}} \right)}^{{W_0}}}}} - \left( {1 - {p_W}} \right)P_{idle}^W} \right)\\
{\left. { \cdot \left( {2{W_0}\frac{{1 - {p_W} - {p_W}{{\left( {2{p_W}} \right)}^{m - 1}}}}{{1 - 2{p_W}}} + 1} \right)} \right]^{ - 1}},
\end{array}
\end{equation}
where ${q_W}$ is the probability that the buffer of the Wi-Fi AP has a packet to transmit, ${p_W}$ is the conditional collision probability given that the Wi-Fi AP attempts transmission, and $P_{idle}^W$ is the probability that the channel is sensed to be idle given that the Wi-Fi AP is not transmitting.  

The probability ${\tau _W}$ that a Wi-Fi AP is attempting transmission in a randomly chosen time slot is obtained in \cite{Malone2007} as follows:
\begin{equation} \label{tauW}
\begin{array}{l}
{\tau _W} = {q_W}P_{idle}^Wb_{{{\left( {0,0} \right)}_e}}^W + \sum\limits_{i \ge 0} {b_{\left( {i,0} \right)}^W} \\
\enspace\quad = b_{{{\left( {0,0} \right)}_e}}^W\left( {\frac{{{q_W}^2{W_0}}}{{\left( {1 - {p_W}} \right)\left( {1 - {q_W}} \right)\left( {1 - {{\left( {1 - {q_W}} \right)}^{{W_0}}}} \right)}} - \frac{{{q_W}^2P_{idle}^W}}{{1 - {q_W}}}} \right).
\end{array}
\end{equation}

\subsection{Cellular Small-Cell Base Station Model}
To access the unlicensed spectrum, cellular SCBSs follow a LBT procedure with the random backoff which is introduced in \cite{3gpp, ericsson}. The cellular SCBSs start monitoring the channel for the duration of time called a clear channel assessment (CCA) period. If the channel is sensed to be idle continuously for the CCA, the backoff mechanism of LBT begins; otherwise, the cellular SCBSs keep monitoring. By setting the length of CCA as the distributed inter-frame space (DIFS) of Wi-Fi, the LBT of cellular SCBSs may seem similar with the carrier sense multiple access with collision avoidance (CSMA/CA) of Wi-Fi. In Wi-Fi APs, when the channel is sensed to be busy, the backoff counter is frozen and reactivated after the channel is sensed to be idle again for DIFS, while the cellular SCBSs restart monitoring the channel for the CCA and then the backoff counter is re-chosen. This is the difference between CSMA/CA and LBT procedure. Also, the LBT has a fixed CW size regardless of a collision unlike the binary exponential backoff of CSMA/CA.

Under the non-saturated traffic condition, the proposed Markov chain for a cellular SCBS consists of a post-backoff stage and a backoff stage similar to the Markov chain of Wi-Fi AP introduced in \cite{Malone2007}. Let ${(k)_e}$ and ${(k)}$ for $k \in \left[ {0,Z-1} \right]$ be the states of post-backoff and backoff stage, respectively, where ${k}$ denotes the backoff counter and ${Z}$ denotes the CW size of the cellular SCBS. At the beginning of LBT procedure, ${k}$ is uniformly chosen in the range of $\left[ {0,Z - 1} \right]$. In state ${(k)_e}$, if the channel is sensed to be busy, the cellular SCBS restarts channel monitoring and then the backoff counter is re-chosen. Otherwise, the transition is occurred to state ${(k-1)_e}$ or ${(k-1)}$ according to whether a packet is arrived or not. When the backoff counter reaches ${0}$, the transmission is attempted. Based on the above descriptions, transition probabilities from state ${(k)_e}$ are given for $k \in [1,Z - 1]$, $l \in \left[ {0,Z - 1} \right]$ and $l \ne k - 1$:
\begin{equation} \label{tr1}
\left\{ \begin{array}{l}
\Pr \left[ {{{\left( {k - 1} \right)}_e}\left| {{{\left( k \right)}_e}} \right.} \right] = \left( {1 - {q_C}} \right)P_{idle}^C + \frac{{\left( {1 - {q_C}} \right)\left( {1 - P_{idle}^C} \right)}}{Z},\\
\Pr \left[ {{{\left( l \right)}_e}\left| {{{\left( k \right)}_e}} \right.} \right] \hspace{18.5pt}= \frac{{\left( {1 - {q_C}} \right)\left( {1 - P_{idle}^C} \right)}}{Z},\,\\
\Pr \left[ {\left( {k - 1} \right)\left| {{{\left( k \right)}_e}} \right.} \right] \hspace{5pt}= {q_C}P_{idle}^C + \frac{{{q_C}\left( {1 - P_{idle}^C} \right)}}{Z},\,\\
\Pr \left[ {\left( l \right)\left| {{{\left( k \right)}_e}} \right.} \right] \hspace{23.3pt}= \frac{{{q_C}\left( {1 - P_{idle}^C} \right)}}{Z},\,
\end{array} \right.
\end{equation}
where ${q_C}$ is the probability that the buffer of the cellular SCBS has packets to transmit, and $P_{idle}^C$ is the probability that the channel is sensed to be idle given that the cellular SCBS is not transmitting.

In state ${(0)_e}$, if an arriving packet is transmitted successfully without a collision, the backoff counter is uniformly chosen in the post-backoff stage. If there is no arriving packet, then the state remains at state ${(0)_e}$. When the buffer of cellular SCBS has a packet, if there is a collision or the channel is sensed to be busy, a transition occurs to one of the states in the backoff stage from state ${(0)_e}$ with the uniformly chosen backoff counter. Based on the above descriptions, the transition probabilities from state ${(0)_e}$ are shown as follows:
\begin{equation} \label{tr2}
\left\{ \begin{array}{l}
\Pr \left[ {{{\left( k \right)}_e}\left| {{{\left( 0 \right)}_e}} \right.} \right] = \frac{{{q_C}P_{idle}^C\left( {1 - {p_C}} \right)}}{Z},k \in \left[ {1,Z - 1} \right],\\
\Pr \left[ {{{\left( 0 \right)}_e}\left| {{{\left( 0 \right)}_e}} \right.} \right] \hspace{0.2pt}= \frac{{{q_C}P_{idle}^C\left( {1 - {p_C}} \right)}}{Z} + \left( {1 - {q_C}} \right),\\
\Pr \left[ {\left( k \right)\left| {{{\left( 0 \right)}_e}} \right.} \right] \hspace{4.2pt}= \frac{{{p_C}{q_C}P_{idle}^C}}{Z} + \frac{{{q_C}\left( {1 - P_{idle}^C} \right)}}{Z},\,k \in \left[ {0,Z - 1} \right],
\end{array} \right.
\end{equation}
where ${p_C}$ is the conditional collision probability given that the cellular SCBS is attempting transmission.

In state ${(k)}$, if the channel is sensed to be busy, the backoff counter is uniformly chosen in the backoff stage; otherwise the backoff counter is decremented by 1. For $k \in \left[ {1, Z - 1} \right]$, the transition probabilities are given as follows:
\begin{equation} \label{tr3}
\left\{ \begin{array}{l}
\Pr \left[ {\left( {k - 1} \right)\left| {\left( k \right)} \right.} \right] = \frac{{{p_C}}}{Z} + (1 - {p_C}),\,\\
\Pr \left[ {\left( l \right)\left| {\left( k \right)} \right.} \right] \hspace{18.3pt} = \frac{{{p_C}}}{Z},l \in [0,Z - 1],\,l \ne k - 1.
\end{array} \right.
\end{equation} 

When the backoff counter reaches ${0}$ in the backoff stage, a packet is transmitted without additional channel sensing \cite{Song2015, Chen2015, ericsson, 3gpp}. If a collision occurs or the buffer has another packet and there is no collision, the backoff counter is uniformly chosen. If the transmission is successful and there is no another packet, the state transition occurs to the post-backoff stage. Therefore, in state ${(0)}$, we have for $k \in \left[ {0, Z - 1} \right]$,
\begin{equation} \label{tr4}
\left\{ \begin{array}{l}
\Pr \left[ {\left( k \right)\left| {\left( 0 \right)} \right.} \right] \hspace{6.5pt}= \frac{{{p_C}}}{Z} + \frac{{{q_C}\left( {1 - {p_C}} \right)}}{Z},\,\\
\Pr \left[ {{{\left( k \right)}_e}\left| {\left( 0 \right)} \right.} \right] = \frac{{\left( {1 - {q_C}} \right)\left( {1 - {p_C}} \right)}}{Z}.
\end{array} \right.
\end{equation}

\textit{Proposition 1:}
Under the given probabilities ${p_C}$, ${q_C}$, $P_{idle}^C$, and ${Z}$, we can derive $b_{{{\left( 0 \right)}_e}}^C$ by using  (\ref{tr1})-(\ref{tr4}) as follows:
\begin{equation}
b_{{{\left( 0 \right)}_e}}^C = {\left[ {\eta \lambda  + P_{idle}^C\left( {1 - {p_C}} \right) + \left( {\eta \mu  + \frac{{\left( {1 - {q_C}} \right)\left( {1 - {p_C}} \right)}}{{{q_C}}}} \right)\gamma } \right]^{ - 1}},
\end{equation}
where $\eta  = \frac{{{p_C}}}{{P_{idle}^C\alpha \left( Z \right)}},$ $\lambda  = \frac{{\left( {1 - \beta \left( 1 \right)} \right){{\left( {P_{idle}^C} \right)}^2}\alpha \left( {Z - 1} \right)}}{{{p_C}}} + \frac{{{{\left( {1 - \beta \left( 1 \right)} \right)}^3}P_{idle}^C\beta \left( {Z - 1} \right)}}{{{q_C}\beta \left( 1 \right)}} + \frac{{{q_C}P_{idle}^C\left[ {{p_C}Z - P_{idle}^C\left( {1 + {p_C}P_{idle}^C} \right)\alpha \left( Z \right)} \right]}}{{p_C^2}},$ $\mu  = \frac{{{{\left( {1 - \beta \left( 1 \right)} \right)}^2}\alpha \left( {Z - 1} \right)}}{{{p_C}{q_C}}} + \frac{{{{\left( {1 - \beta \left( 1 \right)} \right)}^3}\beta \left( {Z - 1} \right)}}{{{q_C}\beta \left( 1 \right)}} + \frac{{P_{idle}^C\left[ {{p_C}Z - P_{idle}^C\left( {1 + {p_C}\left( {1 - {q_C}} \right)} \right)\alpha \left( Z \right)} \right]}}{{p_C^2}},$ $\gamma  = \frac{{{q_C}Z\left[ {\left( {1 - \left( {1 - {p_C}} \right)P_{idle}^C} \right)\left( {1 - \beta \left( 1 \right)} \right)\beta \left( Z \right) - {q_C}Z\beta \left( 1 \right)} \right]}}{{\left( {1 - {p_C}} \right)\left( {1 - {q_C}} \right)\left( {1 - \beta \left( 1 \right)} \right)\beta \left( Z \right)}},$ $\alpha \left( x \right) = 1 - {\left( {1 - {p_C}} \right)^x},$ and $\beta \left( x \right) = 1 - {\left( {\left( {1 - {q_C}} \right)P_{idle}^C} \right)^x}$.

\hspace{14pt}\textit{Proof:} See Appendix A.

A cellular SCBS tries to transmit a packet, if it is in state ${(0)}$ or it is in state ${(0)_e}$ conditioned on the fact that a packet is arrived and the channel is sensed to be idle. Therefore, the probability that a cellular SCBS is attempting transmission can be defined as follows:
\begin{equation} \label{tauC}
{\tau _C} = b_{\left( 0 \right)}^C + {q_C}P_{idle}^Cb_{{{\left( 0 \right)}_e}}^C,
\end{equation}
where $b_{{{\left( 0 \right)}_e}}^C$ and $b_{\left( 0 \right)}^C$ are expressed in terms of ${p_C}$, ${q_C}$, $P_{idle}^C$, and ${Z}$. If the traffic of cellular a SCBS is saturated, i.e., ${q_C} = 1$, the proposed Markov chain and all the equations presented in this work fall back to that of \cite{Song2015}.

\subsection{Channel Idle Probabilities}
Based on the definitions of $P_{idle}^W$ and $P_{idle}^C$ above, if the channel condition is examined at the beginning of each time slot, $P_{idle}^W$ and $P_{idle}^C$ can be simply considered as probabilities that the next slot is empty given that our node is not transmitting. That is, there is no collision conditioned on the fact that our node is transmitting. Therefore, it can be written as follows:
\begin{equation} \label{ch_idle_prob}
\begin{array}{l}
P_{idle}^W = {\left( {1 - {\tau _W}} \right)^{{n_W} - 1}}{\left( {1 - {\tau _C}} \right)^{{n_C}}} = 1 - {p_W},\\
P_{idle}^C = {\left( {1 - {\tau _W}} \right)^{{n_W}}}{\left( {1 - {\tau _C}} \right)^{{n_C} - 1}} = 1 - {p_C}.
\end{array}
\end{equation} 

\subsection{Collision Probability and Throughput}
As \cite{Song2015}, in (\ref{tauW}) and (\ref{tauC}), ${\tau _W}$ and ${\tau _C}$ are functions of ${p_W}$ and ${p_C}$, respectively, and ${p_W}$ and ${p_C}$ can be also expressed by ${\tau _W}$ and ${\tau _C}$ as follows:
\begin{equation} \label{collisionprob}
\begin{array}{l}
{p_W} = 1 - {\left( {1 - {\tau _W}} \right)^{{n_W} - 1}}{\left( {1 - {\tau _C}} \right)^{{n_C}}},\\
{p_C} \,= 1 - {\left( {1 - {\tau _W}} \right)^{{n_W}}}{\left( {1 - {\tau _C}} \right)^{{n_C} - 1}}.
\end{array}
\end{equation}
That is, there is a collision when at least two nodes simultaneously transmit. With ${n_W}$ Wi-Fi APs and ${n_C}$ cellular SCBSs, (\ref{tauW}), (\ref{tauC}), and (\ref{collisionprob}) provide nonlinear simultaneous equations which can be solved numerically for ${\tau_W}$, ${\tau_C}$, ${p_W}$, and ${p_C}$.

Let $P_{tr}^W$ and $P_{tr}^C$ denote the probabilities that there is at least one transmission among Wi-Fi APs and cellular SCBSs, respectively. Since there are ${n_W}$ Wi-Fi APs and ${n_C}$ cellular SCBSs on the same channel in the unlicensed spectrum, $P_{tr}^W$ and $P_{tr}^C$ are represented by ${\tau_W}$ and ${\tau_C}$ as follows:
\begin{equation}
P_{tr}^W = 1 - {\left( {1 - {\tau _W}} \right)^{{n_W}}},\,P_{tr}^C = 1 - {\left( {1 - {\tau _C}} \right)^{{n_C}}}.
\end{equation}
Also, the probabilities $P_s^W$ and $P_s^C$ that exactly one Wi-Fi AP and cellular SCBS among ${n_W}$ Wi-Fi APs and ${n_C}$ cellular SCBSs attempt a transmission under the condition that at least one Wi-Fi and cellular node transmit are represented as follows:
\begin{equation}
\begin{array}{l}
P_s^W = \frac{{{n_W}{\tau _W}{{\left( {1 - {\tau _W}} \right)}^{{n_W} - 1}}}}{{P_{tr}^W}} = \frac{{{n_W}{\tau _W}{{\left( {1 - {\tau _W}} \right)}^{{n_W} - 1}}}}{{1 - {{\left( {1 - {\tau _W}} \right)}^{{n_W}}}}},\\
P_s^C = \frac{{{n_C}{\tau _C}{{\left( {1 - {\tau _C}} \right)}^{{n_C} - 1}}}}{{P_{tr}^C}} = \frac{{{n_C}{\tau _C}{{\left( {1 - {\tau _C}} \right)}^{{n_C} - 1}}}}{{1 - {{\left( {1 - {\tau _C}} \right)}^{{n_C}}}}}.
\end{array}
\end{equation}

Since the elapsed time of each state in the Markov chain is different, we calculate the expected time spent per state. The states are included in the one of several cases: the channel being idle, a successful transmission of Wi-Fi APs or cellular SCBSs, and a collision among Wi-Fi APs or cellular SCBSs or both nodes. Therefore, we consider the probability of each case, and we can express the expected time per state as follows:\begin{equation} \label{T_state}
\begin{array}{l}
{T_{state}}= \left( {1 - P_{tr}^W} \right)\left( {1 - P_{tr}^C} \right)\sigma  + P_{tr}^WP_s^W\left( {1 - P_{tr}^C} \right)T_S^W\\
\hspace{35pt}+ \left( {1 - P_{tr}^W} \right)P_{tr}^CP_s^CT_s^C + P_{tr}^W\left( {1 - P_s^W} \right)\left( {1 - P_{tr}^C} \right)T_c^W\\
\hspace{35pt}+ \left( {1 - P_{tr}^W} \right)P_{tr}^C\left( {1 - P_s^C} \right)T_c^C + P_{tr}^WP_{tr}^CT_c^M,
\end{array}
\end{equation}
where ${\sigma}$ denotes a slot time, and $T_s^W$ and $T_s^C$ denote the expected time taken for a successful transmission of a Wi-Fi AP and a cellular SCBS, respectively. Also, $T_c^W$ and $T_c^C$ denote the expected time of a collision among Wi-Fi APs and among cellular SCBSs, respectively, and ${T_c^M}$ denotes the expected time of a collision among Wi-Fi and cellular nodes. Based on (\ref{T_state}), let ${S_W}$ and ${S_C}$ be the throughputs of Wi-Fi and cellular networks, respectively. Thus, we have
\begin{equation}
{S_W} = \frac{{P_{tr}^WP_s^W\left( {1 - P_{tr}^C} \right){D_W}}}{{{T_{state}}}},{S_C} = \frac{{P_{tr}^CP_s^C\left( {1 - P_{tr}^W} \right){D_C}}}{{{T_{state}}}},
\end{equation}
where ${D_W}$ and ${D_C}$ are the total number of payload bits in the Wi-Fi and cellular networks, respectively. Therefore, the throughputs of Wi-Fi and cellular networks are the function of CW size of cellular SCBSs. Under ${n_W}$ Wi-Fi APs and ${n_C}$ cellular SCBSs, total network throughput ${S_{total}}$ is defined as ${S_{total}} = {n_W}{S_W} + {n_C}{S_C}.$

\subsection{Optimal Contention Window Size}
The optimal CW size of cellular SCBSs is determined as a minimum CW size which maximizes total throughput while satisfying the graceful coexistence. Let ${\bf{q}}$ and ${\bf{n}}$ be a set $\left\{{{q_W},{q_C}} \right\}$ and a set $\left\{{{n_W},{n_C}} \right\}$, respectively. Under the given ${\bf{q}}$ and ${\bf{n}}$, we can formulate an optimization problem to determine the optimal CW size as follows:
\begin{equation} \label{opt}
\begin{array}{l}
{Z^*} = \arg \mathop {\max }\limits_Z {S_{total}}\left( {Z\left| {\bf{q}} \right.,{\bf{n}}} \right)\\
\hspace{14.5pt} = \arg \mathop {\max }\limits_Z {n_W} \cdot S_{co}^W\left( {Z\left| {\bf{q}} \right.} \right) + {n_C} \cdot S_{co}^C\left( {Z\left| {\bf{q}} \right.} \right),
\end{array}
\end{equation}
\hspace{44pt}subject to
\begin{equation*}
\hspace{43pt}\min \left[ {S_{co}^W\left( {Z\left| {\bf{q}} \right.} \right),S_{co}^C\left( {Z\left| {\bf{q}} \right.} \right)} \right] > S_{only}^W\left( {{q_W}} \right),
\end{equation*}
where $S_{co}^W$ and $S_{co}^C$ denote the throughputs of a Wi-Fi AP and a cellular SCBS in the coexistence scenario, respectively, and $S_{only}^W$ denotes the throughput of a Wi-Fi AP where Wi-Fi APs exist only. 
Since the objective function in the problem, i.e., total throughput ${S_{total}}$, cannot be expressed as a closed form, adapting global optimization techniques \cite{horst2013handbook} can be a reasonable approach to find an optimal solution of the problem. While the exact computational complexity depends upon the chosen optimization technique, the proposed analytical approach is clearly less complex than exhaustive simulations by virtue of the computable expression of total throughput.
\section{Numerical Results}
\begin{figure}[h] 
	\centering
	\includegraphics[width=3in]{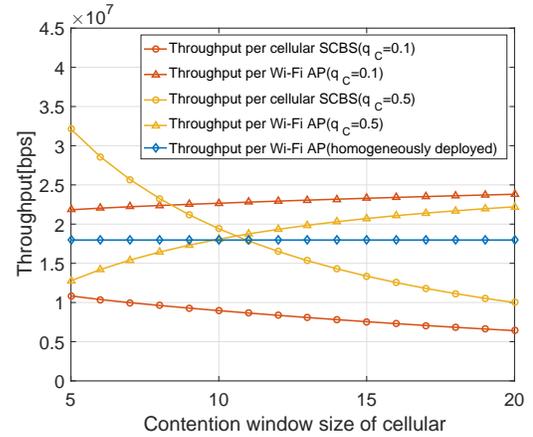}
	\centering
	\caption{Throughputs of Wi-Fi AP and cellular SCBS under $q_W=0.5$ and different $q_C$} \label{searching_optimal_CW}
\end{figure}
In this paper, we consider a scenario with ${n_W=2}$ and ${n_C=1}$. The objective for a simulation is to find the optimal CW size of the cellular SCBSs that maximizes the total throughputs of Wi-Fi and cellular networks while satisfying the graceful coexistence condition. The parameter setting for a simulation are adopted from IEEE 802.11 ac standard as follows: ${D_W} = {D_C} = 12000$ bits, PHY header ${=128}$ bits, MAC header ${=272}$ bits, ACK ${=112}$ bits ${+}$ PHY header, propagation delay $ = 0.1\mu $s, $\sigma = 9\mu $s, SIFS ${=16\mu}$s, DIFS ${=34\mu}$s, ${m=3}$, ${W_0=16}$, and the physical data rate of Wi-Fi APs ${R_W = 100}$ Mbps. Among various global optimization techniques \cite{horst2013handbook}, a linear search method is adopted to obtain the optimal CW size. Through using it, the effect of adjustment in CW size of the cellular SCBS on the coexistence performance can be explicitly identified.  
\begin{figure}[t]
	\begin{minipage}{.48\textwidth}
			\begin{subfigure}[t]{0.48\textwidth}
			\centering
			\includegraphics[width=1.8in]{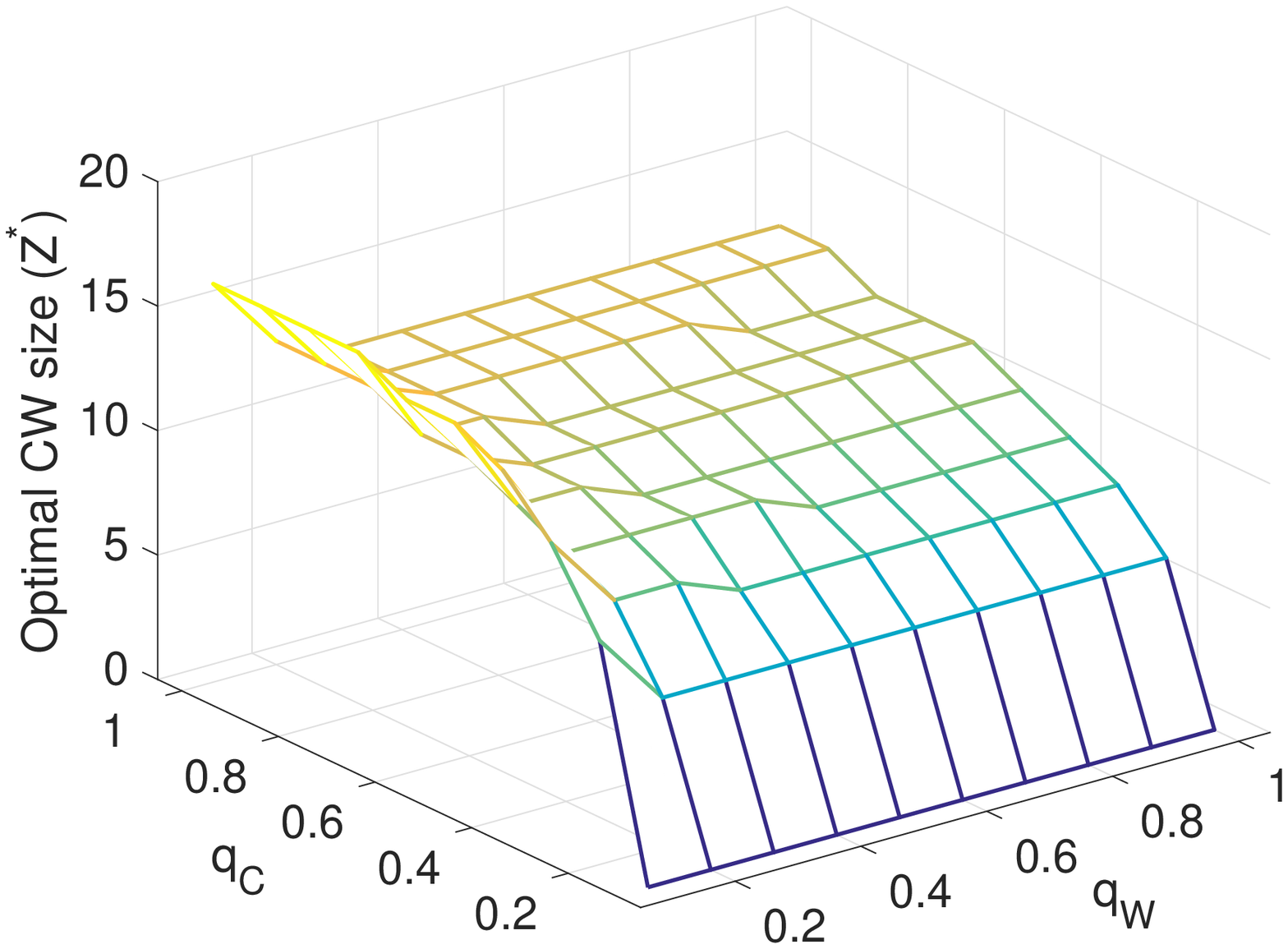}
			\centering
			\caption{}
		\end{subfigure}
		\begin{subfigure}[t]{0.48\textwidth}
			\centering
			\includegraphics[width=1.8in]{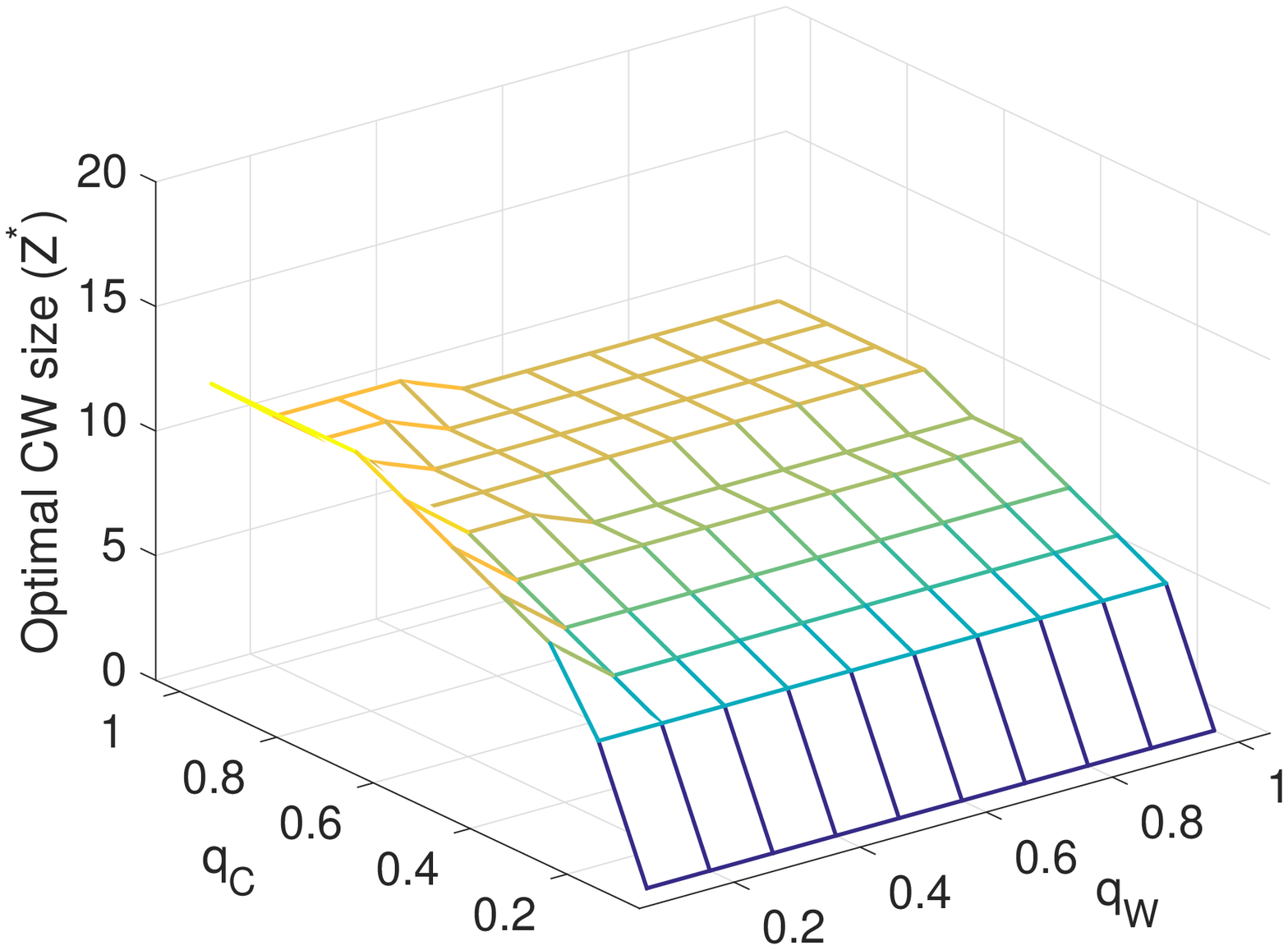}
			\centering
			\caption{}
		\end{subfigure}	\caption{Optimal CW size in all traffic densities of Wi-Fi and cellular networks under (a) ${R_C=100}$ Mbps (b) ${R_C=200}$ Mbps} \label{optimal_CW}
	\end{minipage}\hfill
	\begin{minipage}{.48\textwidth}
		\begin{subfigure}[t]{0.48\textwidth}
			\centering
			\includegraphics[width=1.8in]{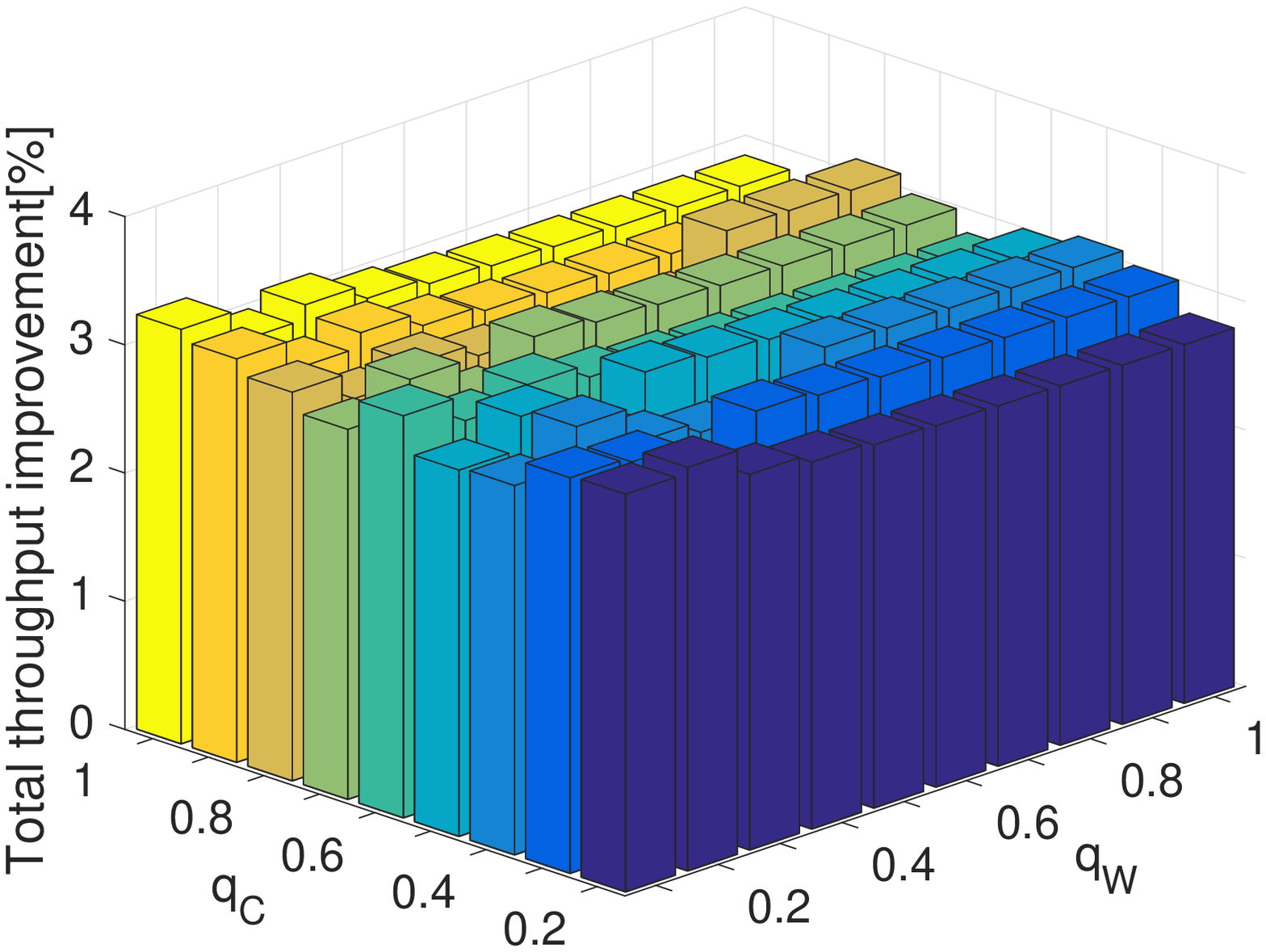}
			\centering
			\caption{}
		\end{subfigure}
		\begin{subfigure}[t]{0.48\textwidth}
			\centering
			\includegraphics[width=1.8in]{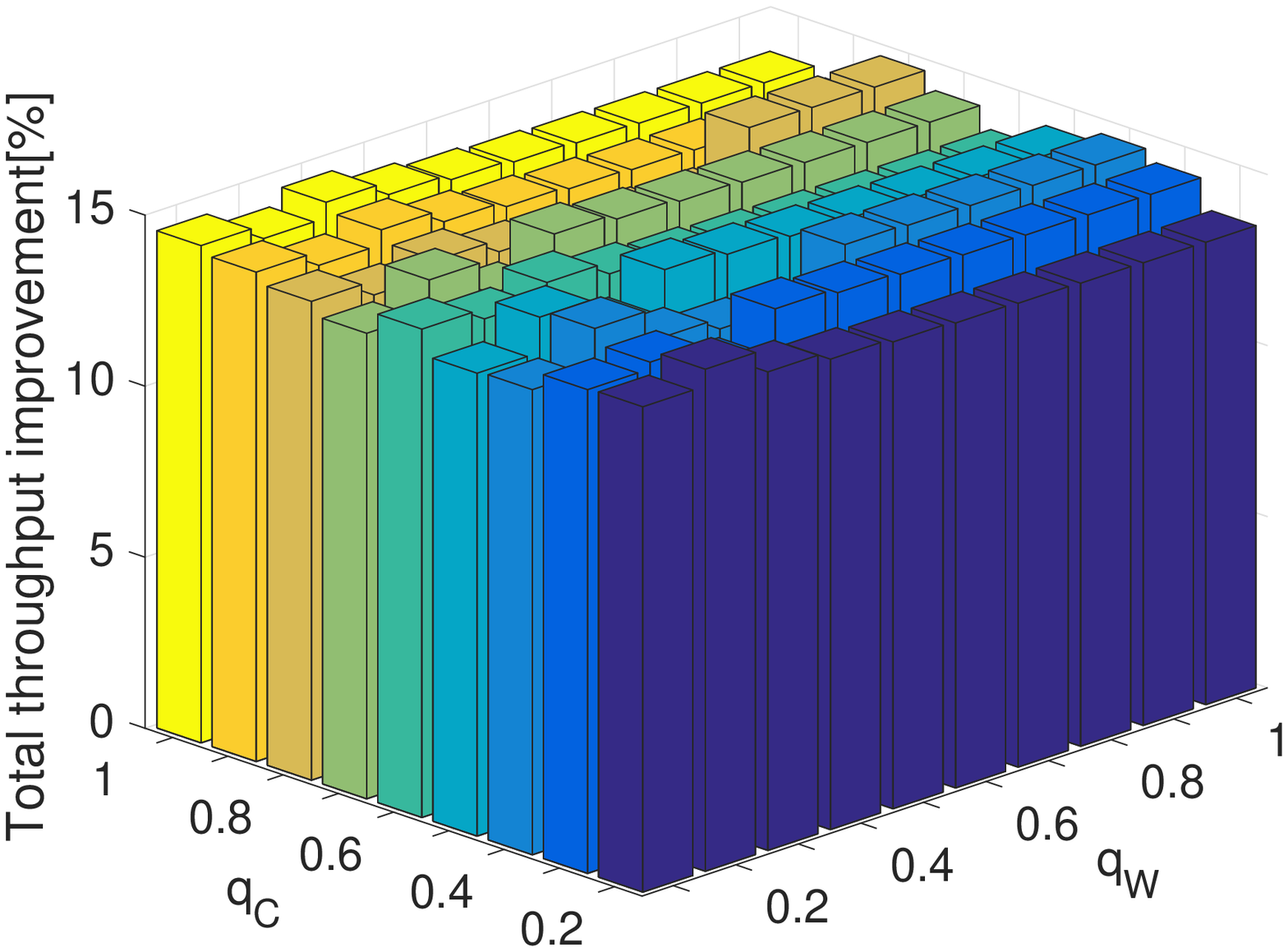}
			\centering
			\caption{}
		\end{subfigure}	\caption{Total throughput improvement in all traffic densities of Wi-Fi and cellular networks with the optimal CW under (a) ${R_C=100}$ Mbps (b) ${R_C=200}$ Mbps} \label{thr_imp}
	\end{minipage}\hfill	
\end{figure}

Fig. \ref{searching_optimal_CW} depicts the throughputs according to the change in the CW size of the cellular SCBS with ${q_W=0.5}$ and different ${q_C}$ values. Note that the cellular SCBS has the same physical data rate with that of the Wi-Fi APs (i.e., ${R_C=100}$ Mbps). Under the case with ${q_C=0.1}$, it is observed that there is no CW size which satisfies the graceful coexistence, because the throughput of the cellular SCBS is lower than the requirement (i.e., graceful coexistence) regardless of the CW size. In case of ${q_C=0.5}$, the throughputs of the Wi-Fi APs and the cellular SCBS are above the requirement under  ${Z=10}$ (i.e., ${Z^*=10}$). From these results, we can say that the optimal CW size does not always exist in all traffic densities of both networks, such that very careful adjustment in the CW size of cellular SCBS is required to satisfy the graceful coexistence.

Fig. \ref{optimal_CW} illustrates the optimal CW size of the cellular SCBS in all traffic densities of Wi-Fi and cellular networks under the different physical data rate of the cellular SCBS. As ${q_C}$ increases, the optimal CW size increases to protect the activities of the Wi-Fi APs and then converges. In addition, the optimal CW size at the low traffic density of the Wi-Fi APs is higher than that at the high traffic density of the Wi-Fi APs, because a small CW size of the cellular SCBS encourages its activity. When the physical data rate of the cellular SCBS increases, the optimal CW size do not need to be high for satisfying the graceful coexistence since the processing time of a packet decreases. Therefore, the optimal CW size can decrease, and it is shown in Figs. \ref{optimal_CW} (a) and (b).  

Fig. \ref{thr_imp} shows total throughput improvement in all traffic densities of both networks under the different physical data rate of the cellular SCBS when the optimal CW size of each case is applied. The growth of physical data rate of the cellular SCBS results in the growth of expected total throughput improvement from 3.0\% to 14.1\%, as shown in Figs. \ref{thr_imp} (a) and (b).

\section{Conclusion}
We investigated the coexistence performance of Wi-Fi and cellular networks sharing the unlicensed spectrum under the non-saturated traffic condition. 
Under the condition, a Markov chain, which describes the behavior of a cellular SCBS with LBT, was proposed, and it was combined with a Markov chain which describes the behavior of a Wi-Fi AP. 
To evaluate the performance, the network scenario with ${n_W}$ Wi-Fi APs and ${n_C}$ cellular SCBSs was considered, and it is compared with the scenario with ${n_W+n_C}$ homogeneously deployed Wi-Fi APs. 
From the numerical results, the optimal CW size of cellular SCBSs was determined in which the total throughput of the network is maximized while satisfying the graceful coexistence condition. 
Our analysis enables the cellular SCBSs to adjust the CW size according to the change in traffic volumes of the networks over time for the graceful coexistence.

\appendices
\section{}
The stationary probabilities of all states can be written in terms of $b_{{{\left( 0 \right)}_e}}^C$. Firstly, $\sum\limits_{k = 0}^{Z - 1} {b_{{{\left( k \right)}_e}}^C}$ can be expressed as follows:
\begin{equation} \label{sum1}
\sum\limits_{k = 0}^{Z - 1} {b_{{{\left( k \right)}_e}}^C}  = P_{idle}^C\left( {1 - {p_C}} \right)b_{{{\left( 0 \right)}_e}}^C + \frac{{\left( {1 - {q_C}} \right)\left( {1 - {p_C}} \right)}}{{{q_C}}}b_{\left( 0 \right)}^C.
\end{equation}
From straightforward recursion, we can also express 
\begin{equation} \label{sum2}
\sum\limits_{k = 0}^{Z - 1} {b_{{{\left( k \right)}_e}}^C}  = \left( {1 - {q_C}} \right)b_{{{\left( 0 \right)}_e}}^C + \sum\limits_{k = 1}^Z {\frac{{1 - {{\left( {1 - \beta \left( 1 \right)} \right)}^k}}}{{\beta \left( 1 \right)}}b_{{{\left( {Z - 1} \right)}_e}}^C} ,
\end{equation}
where $\beta \left( x \right) = 1 - {\left( {\left( {1 - {q_C}} \right)P_{idle}^C} \right)^x}$, and 
$b_{{{\left( {Z - 1} \right)}_e}}^C = \frac{{{q_C}\left( {P_{idle}^C\left( {1 - {p_C}} \right) + 1} \right) - \beta \left( 1 \right)}}{Z}b_{{{\left( 0 \right)}_e}}^C + \frac{{\left( {1 - {q_C}} \right)\left( {1 - {p_C}} \right)}}{Z}b_{\left( 0 \right)}^C + \sum\limits_{k = 0}^{Z - 1} {\frac{{\beta \left( 1 \right) - {q_C}}}{Z}b_{{{\left( k \right)}_e}}^C} $.
After combining (\ref{sum1}) with (\ref{sum2}), $\gamma = b_{\left( 0 \right)}^C/ b_{{{\left( 0 \right)}_e}}^C$ can be derived as 
\begin{equation} \label{ratio}
\gamma  = \frac{{{q_C}Z\left[ {\left( {1 - \left( {1 - {p_C}} \right)P_{idle}^C} \right)\left( {1 - \beta \left( 1 \right)} \right)\beta \left( Z \right) - {q_C}Z\beta \left( 1 \right)} \right]}}{{\left( {1 - {p_C}} \right)\left( {1 - {q_C}} \right)\left( {1 - \beta \left( 1 \right)} \right)\beta \left( Z \right)}},
\end{equation}
In addition, $b_{\left( k \right)}^C$ can be expressed
\begin{equation} \label{b_k}
\begin{array}{l}
b_{\left( k \right)}^C = \frac{{1 - {{\left( {P_{idle}^C} \right)}^{Z - k}}}}{{{p_C}}}b_{\left( {Z - 1} \right)}^C + \frac{{{q_C}P_{idle}^C}}{{\beta \left( 1 \right)}}\\
\hspace{27pt} \cdot \left[ {\frac{{1 - {{\left( {P_{idle}^C} \right)}^{Z - k - 1}}}}{{{p_C}}}} \right.\left. { - \frac{{\beta \left( 1 \right){{\left( {P_{idle}^C} \right)}^{Z - k - 1}} - \beta {{\left( 1 \right)}^{Z - k}}}}{{P_{idle}^C - \beta \left( 1 \right)}}} \right]b_{{{\left( {Z - 1} \right)}_e}}^C,
\end{array}
\end{equation}
where $b_{\left( {Z - 1} \right)}^C = \frac{{{p_C}{q_C}P_{idle}^C}}{Z}b_{{{\left( 0 \right)}_e}}^C + \frac{{{q_C}P_{idle}^C}}{Z}b_{\left( 0 \right)}^C + \sum\limits_{k = 0}^{Z - 1} {\left[ {\frac{{{p_C}{q_C}}}{Z}b_{{{\left( k \right)}_e}}^C + \frac{{{p_C}}}{Z}b_{\left( k \right)}^C} \right]} .$ Then using (\ref{b_k}), $\sum\limits_{k = 0}^{Z - 1} {b_{{{\left( k \right)}_e}}^C} $ can be easily determined, as described in (\ref{b(k)sum}).
\begin{equation} \label{b(k)sum}
\sum\limits_{k = 0}^{Z - 1} {b_{\left( k \right)}^C}  = \eta  \cdot \lambda  \cdot b_{{{\left( 0 \right)}_e}}^C + \eta  \cdot \mu  \cdot b_{\left( 0 \right)}^C,
\end{equation}
where $\eta  = \frac{{{p_C}}}{{P_{idle}^C\alpha \left( Z \right)}},$ $\lambda  = \frac{{\left( {1 - \beta \left( 1 \right)} \right){{\left( {P_{idle}^C} \right)}^2}\alpha \left( {Z - 1} \right)}}{{{p_C}}} + \frac{{{{\left( {1 - \beta \left( 1 \right)} \right)}^3}P_{idle}^C\beta \left( {Z - 1} \right)}}{{{q_C}\beta \left( 1 \right)}} + \frac{{{q_C}P_{idle}^C\left[ {{p_C}Z - P_{idle}^C\left( {1 + {p_C}P_{idle}^C} \right)\alpha \left( Z \right)} \right]}}{{p_C^2}},$ $\mu  = \frac{{{{\left( {1 - \beta \left( 1 \right)} \right)}^2}\alpha \left( {Z - 1} \right)}}{{{p_C}{q_C}}} + \frac{{{{\left( {1 - \beta \left( 1 \right)} \right)}^3}\beta \left( {Z - 1} \right)}}{{{q_C}\beta \left( 1 \right)}} + \frac{{P_{idle}^C\left[ {{p_C}Z - P_{idle}^C\left( {1 + {p_C}\left( {1 - {q_C}} \right)} \right)\alpha \left( Z \right)} \right]}}{{p_C^2}},$ and $\alpha \left( x \right) = 1 - {\left( {1 - {p_C}} \right)^x}.$
With (\ref{sum1}), (\ref{ratio}), (\ref{b(k)sum}), and the normalization condition, $b_{{{\left( 0 \right)}_e}}^C$ is finally expressed as follows:
$\begin{array}{l}
b_{{{\left( 0 \right)}_e}}^C = {\left[ {\eta \lambda  + P_{idle}^C\left( {1 - {p_C}} \right) + \left( {\eta \mu  + \frac{{\left( {1 - {q_C}} \right)\left( {1 - {p_C}} \right)}}{{{q_C}}}} \right)\gamma } \right]^{ - 1}}.
\end{array}$
\ifCLASSOPTIONcaptionsoff
  \newpage
\fi



\bibliographystyle{IEEEtran}
\bibliography{IEEEabrv,refer}
\end{document}